\documentclass[a4paper]{article}

\usepackage{INTERSPEECH2022}
\usepackage{url}
\usepackage{hyperref}
\usepackage{xspace}
\usepackage{verbatim}
\usepackage{multirow}
\usepackage{booktabs}
\usepackage{cite}
\usepackage[table,xcdraw]{xcolor}
\usepackage{array}
\usepackage{tabularx}
\newcolumntype{Y}{>{\centering\arraybackslash}X}

\title{Unified Source-Filter GAN with \\ Harmonic-plus-Noise Source Excitation Generation}
\name{Reo Yoneyama, Yi-Chiao Wu, Tomoki Toda}
\address{Nagoya University, Japan}
\email{yoneyama.reo@g.sp.m.is.nagoya-u.ac.jp, yichiao.wu@g.sp.m.is.nagoya-u.ac.jp, tomoki@icts.nagoya-u.ac.jp}

\begin{document}

\maketitle
\begin{abstract}

This paper introduces a unified source-filter network with a harmonic-plus-noise source excitation generation mechanism.  In our previous work, we proposed unified Source-Filter GAN (uSFGAN) for developing a high-fidelity neural vocoder with flexible voice controllability using a unified source-filter neural network architecture. However, the capability of uSFGAN to model the aperiodic source excitation signal is insufficient, and there is still a gap in sound quality between the natural and generated speech. To improve the source excitation modeling and generated sound quality, a new source excitation generation network separately generating periodic and aperiodic components is proposed. The advanced adversarial training procedure of HiFiGAN is also adopted to replace that of Parallel WaveGAN used in the original uSFGAN. Both objective and subjective evaluation results show that the modified uSFGAN significantly improves the sound quality of the basic uSFGAN while maintaining the voice controllability.
\end{abstract}

\section{Introduction}
Speech generation is a technology for generating speech waveforms on the basis of text or acoustic features. In particular, models that generate waveforms on the basis of acoustic features are called vocoders. In recent years, neural vocoders \cite{wavenet, wavernn, pwg, qpnet, qppwg, lpcnet, nsf, hn-nsf, nhv, periodnet, hn-pwg, hifigan, gan-vocoder} have been extensively studied, and the sound quality of the synthesize speech is almost as good as that of human speech. However, these vocoders have several drawbacks due to their fully data-driven manner. One is the lack of robustness to unknown data. For example, when fundamental frequency ($\text{F}_0$) deviates from the $\text{F}_0$ range of the training data, these neural vocoders often fail to synthesize high-fidelity speech, or the $\text{F}_0$ of the synthesized speech is not consistent with the given $\text{F}_0$. Another shortage is the disability to independently control acoustic parameters, which is essential to the expressive speech generation applications such as entertaining text-to-speech and voice conversion. Consequently, most neural vocoders suffer from lack of the flexible controllability of pitch, timbre, and aperiodicity, which are usually achieved with the conventional parametric vocoders \cite{straight, world}. To address this issue, studies \cite{nsf, hn-nsf, lpcnet, nhv} have attempted to improve the $\text{F}_0$ controllability by introducing the source-filter theory \cite{source_filter} into their deep neural networks. However, the ad hoc constraints and signal processing can cause several problems such as the degradations of sound quality.

Our previous work, unified Source-Filter GAN (uSFGAN) \cite{usfgan}, mimics the source-filter model using a generative adversarial networks (GAN)-based \cite{gan} unified network. The key concepts of uSFGAN are the regularization loss on the output signal of the source-network which requires flat spectral envelopes, the pitch-dependent source excitation signal generation, and the GAN-based training. This approach jointly optimizes the source excitation generation (source-network) and the resonance filtering (filter-network) modules, which outperforms Neural Source-Filter (NSF) \cite{nsf, hn-nsf} and Quasi-Periodic Parallel WaveGAN (QPPWG) \cite{qppwg} in both sound quality and $\text{F}_0$ controllability. However, there is still a gap in sound quality between natural speech and the generated speech of uSFGAN.

In this paper, to bridge the gap, we propose several improvements to uSFGAN. First, we replace the regularization loss with a new one that utilizes the residual spectrum as the target for stable training. Secondly, to tackle the undesired periodicity in the aperiodic components of the source excitation signals caused by the pitch-dependent source-network, we propose a parallel periodic and aperiodic source excitation generation inspired by the success of the harmonic-plus-noise decomposing neural vocoders \cite{nhv, hn-nsf, periodnet, hn-pwg}. Thirdly, we adopt the training procedure of HiFiGAN \cite{hifigan} instead of that of Parallel WaveGAN (PWG) \cite{pwg} to take the $\text{F}_0$ estimation errors into account. According to the objective and subjective evaluation results, the proposed harmonic plus noise uSFGAN (HN-uSFGAN) significantly improves the sound quality while keeping the $\text{F}_0$ controllability.

\section{Baseline uSFGAN}

\begin{figure}[tb]
\begin{center}
    \vspace{-1mm}
    \includegraphics[width=\columnwidth]{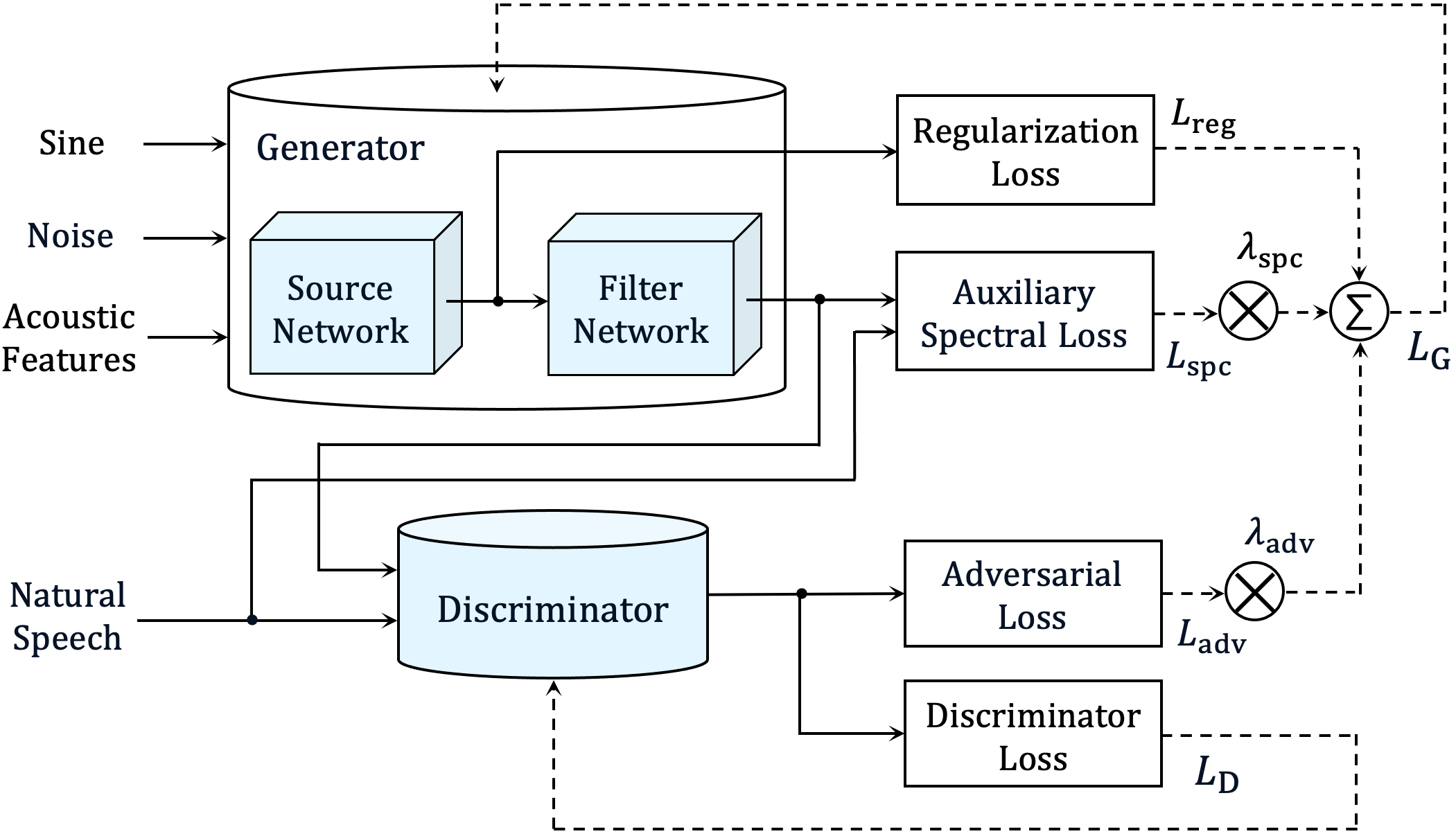}
    \vspace{-5mm}
    \caption{Overall architecture of unified Source-Filter GAN.}
    \label{fig:usfgan}
\end{center}
\vspace{-4mm}
\end{figure}

This section presents the baseline uSFGAN, which mimics a source-filter model using a single unified neural network. Specifically, as shown in Fig. \ref{fig:usfgan}, the overall architecture of the baseline uSFGAN is a general GAN including a neural generator ($G$) and a neural discriminator ($D$). The uSFGAN generator includes source and filter networks, and a regularization loss is applied to the output of the source network for the excitation signal generation. In addition, not only the adversarial losses of the generator and discriminator but also an auxiliary spectral loss is adopted for improving the training stability \cite{pwg}.

\subsection{Unified Source-Filter Networks}

The generator of uSFGAN is factorized into the source excitation generation and resonance filtering networks. The regularization loss on the output signal of the source-network is designed based on the conventional source-filter models \cite{straight, world} which assume that the power of any source excitation signal is constant and its spectral envelopes are flat. Therefore, the regularization loss is formulated as
\begin{equation} \label{math:reg loss}
L_{\text{reg}}(G) = \mathbb{E}_{\bm{z} \sim \mathcal{N}(0, I)} \left[ \frac{1}{N} \lVert \log \hat{E} \rVert_2 \right],
\end{equation}
where $\lVert \cdot \rVert_2$,  $\hat{E}$ and $N$ denote the L2 norm, the magnitude of the source spectral envelopes, and the number of elements in the magnitude, respectively. $\hat{E}$ is extracted using the simplified CheapTrick \cite{cheaptrick} algorithm described in \cite{usfgan}. The baseline uSFGAN jointly optimizes the cascaded source excitation and resonance filtering network to generate reasonable source excitation signals and high-fidelity speech with the source regularization loss and the GAN-based structure. The network architecture is interpretable and tractable with the flexibility of controlling voice components such as $\text{F}_0$.

\subsection{Pitch-dependent Source Excitation Generation}

To capture the long term periodicity of source excitation signals and retain the robustness to unseen $\text{F}_0$ in the inference stage, pitch-dependent dilated convolution neural networks (PDCNNs) \cite{qpnet, qppwg}, which dynamically change the sizes of the receptive fields by multiplying the convolution dilation factors according to the input $\text{F}_0$, are adopted in the source-network. On the other hand, the usual dilated convolution networks (DCNN) with the same structure as PWG \cite{pwg} are adopted in the filter-network. Furthermore, inspired by NSF \cite{nsf}, the inputs of the generator are white noise and the sinusoidal signals corresponding to the input $\text{F}_0$ for efficiently modeling the periodic components of the source excitation signal.

\subsection{PWG-based Adversarial Training}

The training procedure of uSFGAN is based on that of PWG. The PWG discriminator is trained to identify natural samples as $real$ and generated samples as $fake$ by minimizing the following optimization criterion:
\begin{equation}
\begin{split}
L_{D}(G, D) = \mathbb{E}_{\bm{x} \sim p_{\text{data}}} \left[ (1-D(\bm{x}))^2 \right]&
\\ + \mathbb{E}_{\bm{z} \sim \mathcal{N}(0,I)} \left[ D(G(\bm{z}))^2 \right]&,
\end{split}
\end{equation}
where $\bm{x}$ denotes the natural samples, $p_{data}$ denotes the data distribution of the natural samples and $\bm{z}$ is random noise distributed from Gaussian distribution. 
On the other hand, the generator is trained to deceive the discriminator by minimizing the following adversarial loss:
\begin{equation}
L_{\text{adv}}(G, D) = \mathbb{E}_{\bm{z} \sim \mathcal{N}(0,I)} \left[ (1-D(G(\bm{z})))^2 \right].
\end{equation}

GAN-based vocoders often use an auxiliary loss in the spectral-domain to stabilize the training process and encourage the generator to generate speech according to the input auxiliary features. Following PWG, uSFGAN uses the multi-resolution short-time Fourier transform (STFT) loss as the auxiliary spectral loss ($L_{\text{spc}}$). Therefore, the final loss function of the generator can be written as the summation of the regularization loss, the auxiliary spectral loss, and the adversarial loss:
\begin{equation}
L_{G}(G, D) = L_{\text{reg}}(G) + \lambda_{\text{spc}}L_{\text{spc}}(G) + \lambda_{\text{adv}} L_{\text{adv}}(G, D),
\end{equation}
where $\lambda_{\text{spc}}$ and $\lambda_{\text{adv}}$ are loss balancing hyperparameters.

\section{Proposed Harmonic + Noise uSFGAN}

For improving source excitation modeling and generated sound quality, a new residual spectra regularization loss for the generated source excitation signals and a harmonic plus noise source excitation generation are proposed.  We adopt the HiFiGAN-based adversarial training for further speech quality improvement instead of the PWG-based one.

\begin{figure}[tb]
\begin{center}
    \includegraphics[width=\columnwidth]{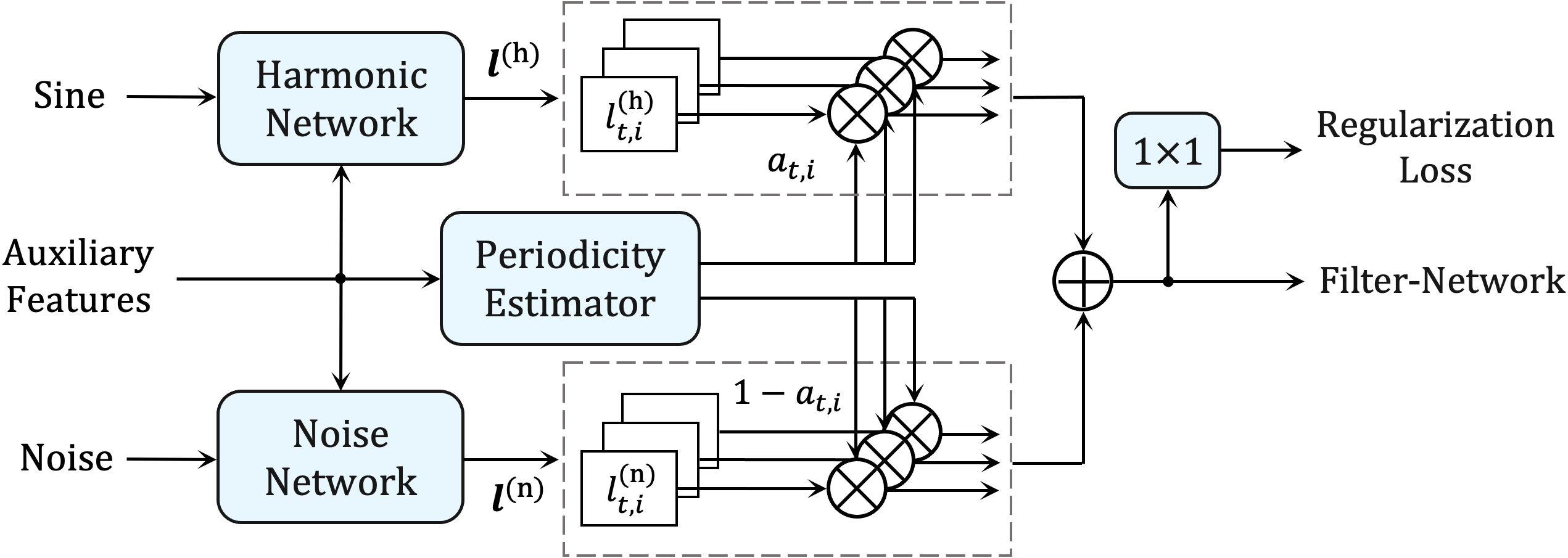}
    \vspace{-5mm}
    \caption{Architecture of proposed harmonic-plus-noise source excitation generation network. $\bigoplus$ denotes element-wise addition.}
    \label{fig:hn-source-network}
\end{center}
\vspace{-8mm}
\end{figure}

\subsection{Residual Spectra Regularization Loss}

According to the source-filter theory \cite{source_filter}, the process of human speech production can be modeled by the source excitation generation (controlling the $\text{F}_0$ and the power) and the resonance filtering (controlling the timbre). While the characteristics of the resonance filtering are represented by the spectral envelopes in the spectral-domain, the characteristics of source excitation generation can be obtained as the residual spectra between the natural speech and the spectral envelopes in the log spectra. Therefore, the residual spectra have reasonable information about source excitation signal.

We propose a new regularization loss utilizing residual spectra as the target of the output source excitation signal. The residual spectra are calculated from the natural speech using the spectral envelopes estimated by CheapTrick \cite{cheaptrick}. The residual spectra regularization loss can be written as
\begin{equation}
    L_{\text{reg}} = \mathbb{E}_{\bm{x} \sim p_{\text{data}}, \bm{z} \sim \mathcal{N}(0, I)} \left[ \frac{1}{N} \lVert \log{f(S)} - \log{f(\hat{S})} \rVert_{1} \right],
\end{equation}
where $\hat{S}$ is the amplitude spectra of the output source excitation signal, $S$ is the amplitude residual spectra which have the same frame-wise average power as that of the target speech, and $N$ is the number of elements in the amplitude spectra. To flexibly model the source excitation while tolerating perceptually acceptable distortion, mel-filter bank $f$ is applied to both amplitude spectra. Unlike the regularization loss of the baseline uSFGAN, the new loss makes the power estimation of the source-network like a real human speech production process. 

\subsection{Harmonic-plus-Noise Source Excitation Generation}

Although PDCNNs are very effective to improve $\text{F}_0$ controllability, they are not suitable for generating unvoiced segments of the source excitation signal. Specifically, the periodic operation sometimes causes undesired periodic components in the unvoiced segments. To improve the modeling of both periodic and aperiodic components of the source excitation signals, we introduce a parallel harmonic-plus-noise source excitation generation architecture as shown in Fig. \ref{fig:hn-source-network}. 

The harmonic-plus-noise source-network consists of three modules: harmonic-network, noise-network, and periodicity-estimator. The harmonic-network is composed of several PDCNN-based blocks, and the latent feature $\bm{l}^{\text{(h)}}$ corresponding to the periodic components of the source excitation signal is generated on the basis of the input sinusoidal signal and auxiliary features. On the other hand, the noise-network is composed of several CNN-based blocks, and the latent feature $\bm{l}^{\text{(n)}}$ corresponding to the aperiodic components of the source excitation signal is generated on the basis of the input random noise signal and auxiliary features.

The two latent features are element-wise summed using the channel-wise and sample-wise weights $\bm{a}$. These weights are corresponding to the speech periodicity and estimated by the periodicity-estimator according to the input auxiliary features. The source excitation latent feature $\bm{l}$ is formulated as
\begin{equation}
    l_{t, i} = a_{t, i} \cdot l_{t, i}^{\text{(h)}} + (1 - a_{t, i}) \cdot l_{t, i}^{\text{(n)}},
\end{equation}
where the subscripts indicate the $i^{\text{th}}$ channel of the $t^{\text{th}}$ sample of each latent feature or the weights. After the harmonic-plus-noise excitation generation, $\bm{l}$ is passed to the filter-network to generate the synthesized speech and also to the $1 \times 1$ convolution to generate the source excitation signal by reducing the number of channels to one for the regularization loss calculation. We set the the number of channels in $\bm{l}$ to 64 as the same number of the residual channel.

This architecture is inspired by multi-band harmonic-plus-noise PWG (HN-PWG) \cite{hn-pwg} in which periodic and aperiodic output signals through band-pass-filters are summed up with estimated energy for each subband. However, our approach applies harmonic-plus-noise decomposition to the source excitation generation while HN-PWG applies it to the final speech generation. Moreover, we don't adopt band-pass-filtering but estimate channel-wise energy in latent space, promoting the network decomposing periodic and aperiodic components in more flexible way.

\subsection{HiFiGAN-based Adversarial Training}

Inspired by the work of You et al. \cite{gan-vocoder} showing that HiFiGAN's multi-period and multi-scale discriminators \cite{hifigan} are very effective to improve the sound quality of neural vocoders, the effectiveness of these discriminators for our proposed method is explored. Moreover, the auxiliary mel-spectral loss from HiFiGAN is also adopted. Specifically, since our method depends on an external $\text{F}_0$ extractor, the error propagations from the feature extraction errors are inevitable, and these errors usually cause significant degradation of the training losses, such as the multi-resolution STFT loss in PWG and the feature matching loss in HiFiGAN. Consequently, even if the generator can generate sufficiently natural speech according to the input $\text{F}_0$, the inevitable mismatches of $\text{F}_0$s and phases still lead to large multi-resolution STFT and feature matching losses. On the other hand, the mel-spectral loss has a relatively large tolerance for those mismatches because the resolution of each frequency band is reduced according to human perception. Since the generator can still learn the phase information via the adversarial training mechanism, the multi-resolution STFT loss is not necessary for the proposed model. Although the feature matching loss is adopted in HiFiGAN, the proposed method excludes it to avoid the mismatch issues.

\section{Experimental Evaluations}

\subsection{Experimental Setting}

\subsubsection{Data Preparation}

In the following experiments, we used the VCTK corpus \cite{vctk} composed of 110 speakers, and the sampling frequency was set to 24000 Hz. We limited $\text{F}_0$ range of training data from 70 Hz to 340 Hz and excluded two speakers (p271, p300) from training data to use for evaluation. Consequently, we used 99 speakers as seen speakers and the rest as unseen speakers. The testing data consisted of 100 randomly selected utterances from the seen speakers and 100 randomly selected utterances from the unseen speakers. The total number of the training utterances was 35671, which was around 33 hours.

\subsubsection{Model Details}

We compared proposed HN-uSFGAN with two baselines and four ablation models as follows:
\begin{itemize}
\item $\textbf{WORLD}$: Conventional source-filter vocoder \cite{world}.
\item $\textbf{uSFGAN}$: Baseline uSFGAN vocoder \cite{usfgan}. $\lambda_{\text{spc}}$ and $\lambda_{\text{adv}}$ were set to 1.0 and 4.0, respectively.
\item $\textbf{HN-uSFGAN}$: Proposed model with the harmonic-network (20 PDCNN layers with 4 dilation cycles), the noise-network (5 CNN layers), and the filter-network (30 DCNN layers with 3 dilation cycles). $\lambda_{\text{spc}}$ and $\lambda_{\text{adv}}$ were set to 15.0 and 1.0, respectively.
\item $\textbf{- Reg-loss}$: Proposed model trained without the residual spectra loss.
\item $\textbf{- HN-SN}$: Proposed model without the harmonic-plus-noise source-network but with the source-network of baseline uSFGAN.
\item $\textbf{- HiFi-D}$: Proposed model without the HiFiGAN discriminators \cite{hifigan} but with the PWG one.
\item $\textbf{- Mel-loss}$: Proposed model trained without the mel-spectra loss but with the multi-resolution STFT loss \cite{pwg}.
\end{itemize}
The baseline uSFGAN was trained with only the auxiliary losses for the first 100k iterations and trained with the discriminator in the rest 500k steps using the RAdam optimizer \cite{radam}. On the other hand, HN-uSFGAN followed the HiFiGAN training procedure to simultaneously train the generator and the discriminators from scratch for 600k iterations using the Adam optimizer \cite{adam}.

The acoustic features extracted by WORLD with a 5 ms shift length were adopted for all neural vocoders. Specifically, we used one-dimensional continuous $\text{F}_0$, 41 dimensional mel-cepstral coefficients, and three-dimensional coded aperiodicity.

\subsection{Experimental Results}

We conducted both objective and subjective evaluations. For the objective evaluation, three measurements were used: root mean square error of log $\text{F}_0$: RMSE [Hz], voiced or unvoiced decision error: $\text{V/UV}$ [\%], and mel-cepstral distortion: MCD [dB]. For the subjective evaluation, we conducted opinion tests on sound quality with 10 subjects, and each subject evaluated 20 utterances per method and condition.\footnote{We provide audio samples at \url{https://chomeyama.github.io/HN-UnifiedSourceFilterGAN-Demo/}.}

\begin{table}
\vspace{-4mm}
\caption{Results of objective and subjective evaluations. The best results and those that are not significantly different from them are highlighted in bold.}
\label{table:results}
\fontsize{8pt}{10pt}
\selectfont
{%
\begin{tabularx}{\columnwidth}{XYYYYY}
\toprule
\multicolumn{1}{c}{method} & \multicolumn{1}{c}{RMSE $\downarrow$} & \multicolumn{1}{c}{V/UV $\downarrow$} & \multicolumn{1}{c}{MCD $\downarrow$} & \multicolumn{2}{c}{MOS $\uparrow$} \\ 
\midrule
\multicolumn{1}{c}{} & \multicolumn{5}{c}{\cellcolor[HTML]{EFEFEF}Reconstruction (original $\text{F}_0$)} \\
\multicolumn{1}{l}{Natural} & -- & -- & -- & \multicolumn{2}{c}{$\underline{3.90 \pm 0.04}$} \\
\multicolumn{1}{l}{WORLD} & $\textbf{0.05}$ & 12 & 3.51 & \multicolumn{2}{c}{$3.64 \pm 0.06$} \\
\multicolumn{1}{l}{uSFGAN} & $\textbf{0.05}$ & 12 & 3.09 & \multicolumn{2}{c}{$3.66 \pm 0.05$} \\
\multicolumn{1}{l}{HN-uSFGAN} & $\textbf{0.05}$ & 9 & 2.82 & \multicolumn{2}{c}{$\textbf{3.79} \pm 0.05$} \\
\multicolumn{1}{l}{~ - Reg-loss} & $\textbf{0.05}$ & 10 & 2.95 & \multicolumn{2}{c}{$3.66 \pm 0.05$} \\
\multicolumn{1}{l}{~ - HN-SN} & 0.06 & 11 & 3.22 & \multicolumn{2}{c}{$\textbf{3.80} \pm 0.05$} \\
\multicolumn{1}{l}{~ - HiFi-D} & $\textbf{0.05}$ & $\textbf{8}$ & $\textbf{2.65}$ & \multicolumn{2}{c}{$3.56 \pm 0.05$} \\
\multicolumn{1}{l}{~ - Mel-loss} & 0.06 & 11 & 3.04 & \multicolumn{2}{c}{$\textbf{3.76} \pm 0.05$} \\
\midrule
\multicolumn{1}{c}{} & \multicolumn{5}{c}{\cellcolor[HTML]{EFEFEF}$\text{F}_0$ transformation ($2^{-0.75} \times \text{F}_0$)} \\
\multicolumn{1}{l}{WORLD} & $\textbf{0.08}$ & 16 & 3.62 & \multicolumn{2}{c}{$3.60 \pm 0.07$} \\
\multicolumn{1}{l}{uSFGAN} & 0.11 & 14 & 3.33 & \multicolumn{2}{c}{$3.47 \pm 0.07$} \\
\multicolumn{1}{l}{HN-uSFGAN} & 0.11 & 14 & 3.08 & \multicolumn{2}{c}{$\textbf{3.76} \pm 0.06$} \\
\multicolumn{1}{l}{~ - Reg-loss} & 0.13 & 18 & 3.20 & \multicolumn{2}{c}{$\textbf{3.70} \pm 0.06$} \\
\multicolumn{1}{l}{~ - HN-SN} & 0.13 & 17 & 3.34 & \multicolumn{2}{c}{$\textbf{3.80} \pm 0.06$} \\
\multicolumn{1}{l}{~ - HiFi-D} & 0.09 & $\textbf{11}$ & $\textbf{2.90}$ & \multicolumn{2}{c}{$3.62 \pm 0.06$} \\
\multicolumn{1}{l}{~ - Mel-loss} & 0.15 & 17 & 3.43 & \multicolumn{2}{c}{$3.67 \pm 0.06$} \\
\midrule
\multicolumn{1}{c}{} & \multicolumn{5}{c}{\cellcolor[HTML]{EFEFEF}$\text{F}_0$ transformation ($2^{0.75} \times \text{F}_0$)} \\
\multicolumn{1}{l}{WORLD} & 0.09 & $\textbf{12}$ & 3.92 & \multicolumn{2}{c}{$\textbf{3.39} \pm 0.06$} \\
\multicolumn{1}{l}{uSFGAN} & $\textbf{0.08}$ & 17 & 3.62 & \multicolumn{2}{c}{$3.30 \pm 0.06$} \\
\multicolumn{1}{l}{HN-uSFGAN} & 0.10 & $\textbf{12}$ & 3.37 & \multicolumn{2}{c}{$\textbf{3.48} \pm 0.06$} \\
\multicolumn{1}{l}{~ - Reg-loss} & 0.11 & 16 & 3.45 & \multicolumn{2}{c}{$\textbf{3.36} \pm 0.07$} \\
\multicolumn{1}{l}{~ - HN-SN} & 0.10 & $\textbf{12}$ & 3.80 & \multicolumn{2}{c}{$\textbf{3.45} \pm 0.06$} \\
\multicolumn{1}{l}{~ - HiFi-D} & 0.10 & 13 & $\textbf{3.32}$ & \multicolumn{2}{c}{$3.28 \pm 0.07$} \\
\multicolumn{1}{l}{~ - Mel-loss} & 0.10 & 13 & 3.39 & \multicolumn{2}{c}{$\textbf{3.42} \pm 0.06$} \\
\bottomrule
\end{tabularx}%
}
\vspace{-5mm}
\end{table}

\subsubsection{Speech Reconstruction}

First, we evaluated the performance of speech reconstruction (i.e., original $\text{F}_0$) and the evaluation results are shown in the top rows of Table\ref{table:results}. Since there was not significant difference between seen and unseen speakers, we have combined all the results together. The results show that HN-uSFGAN outperforms the baselines in all objective metrics and perceptual sound quality. According to the ablation results, we can find that the proposed method achieves the top performances in both the objective and subjective evaluations, which demonstrates the effectiveness of each proposed module especially the residual spectral loss and the HiFiGAN discriminators for perceptual quality improvements.

\subsubsection{Speech Generation with Transformed $F_0$}

Next, we evaluated the performances of $\text{F}_0$ transformation with factors of $2^{-0.75} \simeq 0.59$ and $2^{0.75} \simeq 1.68$. The ground truth $\text{F}_0$s were determined by multiplying the $\text{F}_0$s extracted from natural speech with the scale factors, and they were also adopted as the auxiliary $\text{F}_0$ for all models. The evaluation results are shown in the middle and bottom rows of Table\ref{table:results}. According to the results, we can find that the proposed HN-uSFGAN generates speech with better V/UV error rates, MCDs, and MOS scores for both $\text{F}_0$ factors than WORLD while maintaining almost the same $\text{F}_0$ accuracy as WORLD. Again, these results show the effectiveness of the proposed model to generate high-fidelity speech while attaining the high $\text{F}_0$ controllability. Since flexibly manipulating $\text{F}_0$ is an essential feature of conventional vocoders, the proposed HN-uSFGAN vocoder meets the definition of a vocoder better than the other neural vocoders. In addition, the baseline uSFGAN significantly inferior to other models in sound quality. We attribute the degradation to its capability to deal with multiple speakers and unseen speakers.

\subsection{Visualization of Source Excitation Signals}

\begin{figure}[!t]
\vspace{-5mm}
\begin{center}
    \includegraphics[width=0.92\columnwidth]{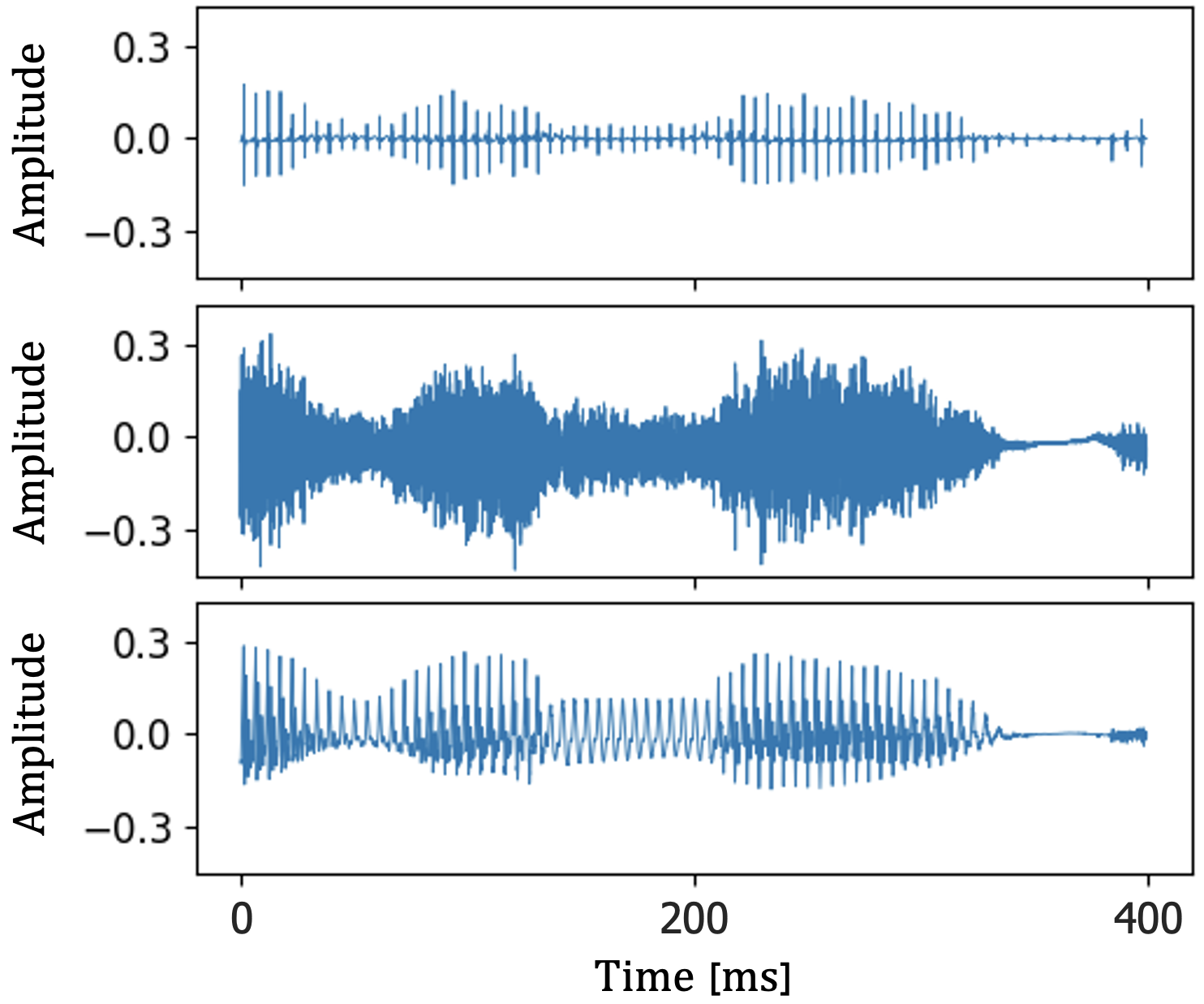}
    \vspace{-2mm}
    \caption{Waveforms of output periodic and aperiodic source excitation signals and the natural speech from top to bottom.}
    \label{fig:har_noi_ref}
    \vspace{2mm}
    \includegraphics[width=0.92\columnwidth]{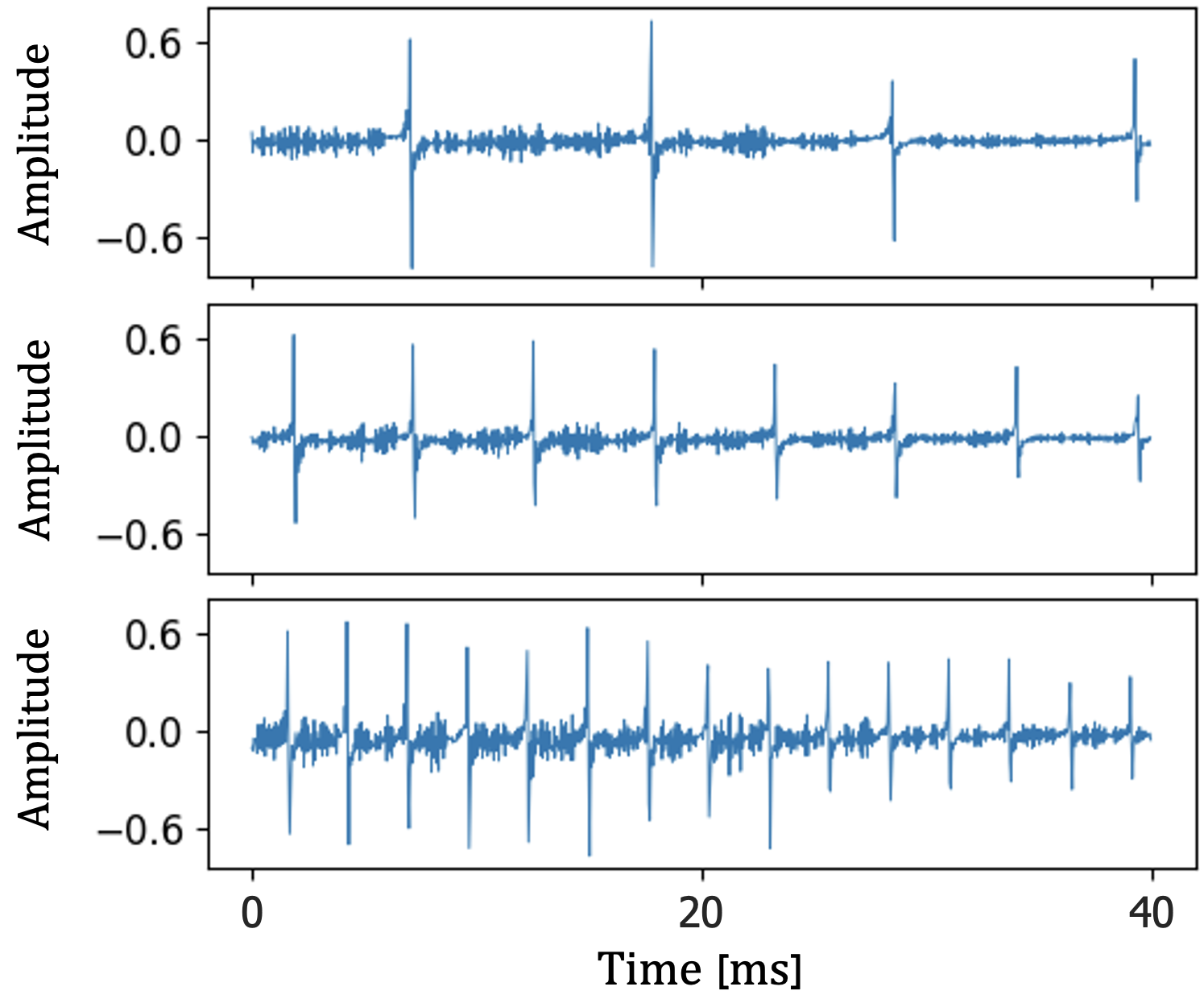}
    \vspace{-2mm}
    \caption{Waveforms of output source excitation signals with different $\text{F}_0$ scaling factors (0.5, 1.0, and 2.0 from top to bottom). Original $\text{F}_0$ values over this segment are around 180 Hz.}
    \label{fig:source_wave_f0_trans}
\end{center}
\vspace{-10mm}
\end{figure}

To investigate the behaviors of proposed HN-uSFGAN, we visualized its output source excitation signals. As shown in Fig. \ref{fig:har_noi_ref}, the generated periodic and aperiodic source excitation signals demonstrate the high capability of HN-uSFGAN to separately model the periodic and aperiodic components. Moreover, Fig. \ref{fig:source_wave_f0_trans} shows that HN-uSFGAN accurately reconstructs the source excitation signals according to the input $\text{F}_0$ and achieves a high capability to handle both the periodic and aperiodic components. 

\section{Conclusions}

The proposed HN-uSFGAN has improved both the speech quality and $\text{F}_0$ controllability of the baseline uSFGAN by harmonic-plus-noise source excitation generation mechanism and the HiFiGAN-based adversarial training procedure. The proposed method has also shown comparative $\text{F}_0$ controllability as the WORLD vocoder which is a conventional vocoder with strong $\text{F}_0$ controllability while significantly outperforming WORLD in sound quality. For future work, we intend to further improve the sound quality of HN-uSFGAN to tackle the occasional buzzy noises in the unvoiced segments.

\noindent\textbf{Acknowledgement}: This work was supported in part by JST, CREST Grant Number JPMJCR19A3 and JSPS KAKENHI Grant Number 21H05054.

\bibliographystyle{IEEEtran}

\bibliography{hnusfgan}

\end{document}